\documentstyle[12pt,psfig]{article}

\catcode`\@=11
\long\def\@makefntext#1{ 
\protect\noindent \hbox to 3.2pt {\hskip-.9pt  
$^{{\ninerm\@thefnmark}}$\hfil}#1\hfill} 

\def\thefootnote{\fnsymbol{footnote}}
 \def\@makefnmark{\hbox to 0pt{$^{\@thefnmark}$\hss}}  
	
\def\ps@myheadings{\let\@mkboth\@gobbletwo
\def\@oddhead{\hbox{} 
\rightmark\hfil\ninerm\thepage}   
\def\@oddfoot{}\def\@evenhead{\ninerm\thepage\hfil 
\leftmark\hbox{}}\def\@evenfoot{}
\def\sectionmark##1{}\def\subsectionmark##1{}}

\begin{document}

\newcommand{\gsim}{\mbox{ \raisebox{-1.0ex}{$\stackrel{\textstyle >}
{\textstyle \sim}$ }}}
\newcommand{\lsim}{\mbox{ \raisebox{-1.0ex}{$\stackrel{\textstyle <}
{\textstyle \sim}$ }}}

\newcommand{\symbolfootnote}{\renewcommand{\thefootnote}
	{\fnsymbol{footnote}}}
\renewcommand{\thefootnote}{\fnsymbol{footnote}}
\newcommand{\alphfootnote}
	{\setcounter{footnote}{0}
	 \renewcommand{\thefootnote}{\sevenrm\alph{footnote}}}

\newcounter{sectionc}\newcounter{subsectionc}\newcounter{subsubsectionc}
\renewcommand{\section}[1] {\vspace{0.6cm}\addtocounter{sectionc}{1} 
\setcounter{subsectionc}{0}\setcounter{subsubsectionc}{0}\noindent 
	{\bf\thesectionc. #1}\par\vspace{0.4cm}}
\renewcommand{\subsection}[1] {\vspace{0.6cm}\addtocounter{subsectionc}{1} 
	\setcounter{subsubsectionc}{0}\noindent 
	{\it\thesectionc.\thesubsectionc. #1}\par\vspace{0.4cm}}
\renewcommand{\subsubsection}[1] {\vspace{0.6cm}\addtocounter{subsubsectionc}{1}
	\noindent {\rm\thesectionc.\thesubsectionc.\thesubsubsectionc. 
	#1}\par\vspace{0.4cm}}
\newcommand{\nonumsection}[1] {\vspace{0.6cm}\noindent{\bf #1}
	\par\vspace{0.4cm}}
					         
\newcounter{appendixc}
\newcounter{subappendixc}[appendixc]
\newcounter{subsubappendixc}[subappendixc]
\renewcommand{\thesubappendixc}{\Alph{appendixc}.\arabic{subappendixc}}
\renewcommand{\thesubsubappendixc}
	{\Alph{appendixc}.\arabic{subappendixc}.\arabic{subsubappendixc}}

\renewcommand{\appendix}[1] {\vspace{0.6cm}
        \refstepcounter{appendixc}
        \setcounter{figure}{0}
        \setcounter{table}{0}
        \setcounter{equation}{0}
        \renewcommand{\thefigure}{\Alph{appendixc}.\arabic{figure}}
        \renewcommand{\thetable}{\Alph{appendixc}.\arabic{table}}
        \renewcommand{\theappendixc}{\Alph{appendixc}}
        \renewcommand{\theequation}{\Alph{appendixc}.\arabic{equation}}
        \noindent{\bf Appendix \theappendixc #1}\par\vspace{0.4cm}}
\newcommand{\subappendix}[1] {\vspace{0.6cm}
        \refstepcounter{subappendixc}
        \noindent{\bf Appendix \thesubappendixc. #1}\par\vspace{0.4cm}}
\newcommand{\subsubappendix}[1] {\vspace{0.6cm}
        \refstepcounter{subsubappendixc}
        \noindent{\it Appendix \thesubsubappendixc. #1}
	\par\vspace{0.4cm}}

\def\abstracts#1{{
	\centering{\begin{minipage}{30pc}\tenrm\baselineskip=12pt\noindent
	\centerline{\tenrm ABSTRACT}\vspace{0.3cm}
	\parindent=0pt #1
	\end{minipage} }\par}} 

\newcommand{\bibit}{\it}
\newcommand{\bibbf}{\bf}
\renewenvironment{thebibliography}[1]
	{\begin{list}{\arabic{enumi}.}
	{\usecounter{enumi}\setlength{\parsep}{0pt}
\setlength{\leftmargin 1.25cm}{\rightmargin 0pt}
	 \setlength{\itemsep}{0pt} \settowidth
	{\labelwidth}{#1.}\sloppy}}{\end{list}}

\topsep=0in\parsep=0in\itemsep=0in
\parindent=1.5pc

\newcounter{itemlistc}
\newcounter{romanlistc}
\newcounter{alphlistc}
\newcounter{arabiclistc}
\newenvironment{itemlist}
    	{\setcounter{itemlistc}{0}
	 \begin{list}{$\bullet$}
	{\usecounter{itemlistc}
	 \setlength{\parsep}{0pt}
	 \setlength{\itemsep}{0pt}}}{\end{list}}

\newenvironment{romanlist}
	{\setcounter{romanlistc}{0}
	 \begin{list}{$($\roman{romanlistc}$)$}
	{\usecounter{romanlistc}
	 \setlength{\parsep}{0pt}
	 \setlength{\itemsep}{0pt}}}{\end{list}}

\newenvironment{alphlist}
	{\setcounter{alphlistc}{0}
	 \begin{list}{$($\alph{alphlistc}$)$}
	{\usecounter{alphlistc}
	 \setlength{\parsep}{0pt}
	 \setlength{\itemsep}{0pt}}}{\end{list}}

\newenvironment{arabiclist}
	{\setcounter{arabiclistc}{0}
	 \begin{list}{\arabic{arabiclistc}}
	{\usecounter{arabiclistc}
	 \setlength{\parsep}{0pt}
	 \setlength{\itemsep}{0pt}}}{\end{list}}

\newcommand{\fcaption}[1]{
        \refstepcounter{figure}
        \setbox\@tempboxa = \hbox{\tenrm Fig.~\thefigure. #1}
        \ifdim \wd\@tempboxa > 6in
           {\begin{center}
        \parbox{6in}{\tenrm\baselineskip=12pt Fig.~\thefigure. #1 }
            \end{center}}
        \else
             {\begin{center}
             {\tenrm Fig.~\thefigure. #1}
              \end{center}}
        \fi}

\newcommand{\tcaption}[1]{
        \refstepcounter{table}
        \setbox\@tempboxa = \hbox{\tenrm Table~\thetable. #1}
        \ifdim \wd\@tempboxa > 6in
           {\begin{center}
        \parbox{6in}{\tenrm\baselineskip=12pt Table~\thetable. #1 }
            \end{center}}
        \else
             {\begin{center}
             {\tenrm Table~\thetable. #1}
              \end{center}}
        \fi}

\def\@citex[#1]#2{\if@filesw\immediate\write\@auxout
	{\string\citation{#2}}\fi
\def\@citea{}\@cite{\@for\@citeb:=#2\do
	{\@citea\def\@citea{,}\@ifundefined
	{b@\@citeb}{{\bf ?}\@warning
	{Citation `\@citeb' on page \thepage \space undefined}}
	{\csname b@\@citeb\endcsname}}}{#1}}

\newif\if@cghi
\def\cite{\@cghitrue\@ifnextchar [{\@tempswatrue
	\@citex}{\@tempswafalse\@citex[]}}
\def\citelow{\@cghifalse\@ifnextchar [{\@tempswatrue
	\@citex}{\@tempswafalse\@citex[]}}
\def\@cite#1#2{{$\null^{#1}$\if@tempswa\typeout
	{IJCGA warning: optional citation argument 
	ignored: `#2'} \fi}}
\newcommand{\citeup}{\cite}

\def\fnm#1{$^{\mbox{\scriptsize #1}}$}
\def\fnt#1#2{\footnotetext{\kern-.3em
	{$^{\mbox{\sevenrm #1}}$}{#2}}}

\font\twelvebf=cmbx10 scaled\magstep 1
\font\twelverm=cmr10 scaled\magstep 1
\font\twelveit=cmti10 scaled\magstep 1
\font\elevenbfit=cmbxti10 scaled\magstephalf
\font\elevenbf=cmbx10 scaled\magstephalf
\font\elevenrm=cmr10 scaled\magstephalf
\font\elevenit=cmti10 scaled\magstephalf
\font\bfit=cmbxti10
\font\tenbf=cmbx10
\font\tenrm=cmr10
\font\tenit=cmti10
\font\ninebf=cmbx9
\font\ninerm=cmr9
\font\nineit=cmti9
\font\eightbf=cmbx8
\font\eightrm=cmr8
\font\eightit=cmti8

\begin{flushright}
  \begin{tabular}[t]{l} 
  KEK-TH-564\\
  March 1998
 \end{tabular}
 \end{flushright}
\vspace*{0.5cm}
\centerline{\tenbf  Kaon and Muon Decay Phenomenology
\footnote{
    Talk presented
    at International KEK Workshop on Kaon, Muon, Neutrino Physics
    and Future, October 31 - November 1, 1997, KEK, Tsukuba, Japan.
}
}
\vspace{0.8cm}
\centerline{\tenrm YASUHIRO OKADA}
\baselineskip=13pt
\centerline{\tenit KEK}
\baselineskip=12pt
\centerline{\tenit Oho 1-1, Tsukuba 305, Japan}
\vspace{0.3cm}
\vspace{0.9cm}
\abstracts{Theoretical aspects of kaon and muon decays are reviewed.
In particular, three topics on kaon and muon decays,
namely $K\rightarrow \pi\nu \bar{\nu}$ processes, T violation in
$K^+ \rightarrow \pi^0\mu^+ \nu$ decay and lepton flavor
violation in muon decay are considered. Theoretical backgrounds and possible
significance on new physics searches by these processes are
discussed.}

\vfil
\twelverm   
\baselineskip=14pt

\section{Introduction}

Current view of the elementary particle physics is based on
SU(3) $\times$ SU(2) $\times$ U(1)
gauge theory of quarks and leptons.  This theory called the 
Standard Model (SM) has been intensively studied in various 
processes.  In order to establish the SM and search for
physics beyond the SM two directions are considered 
in experiments of high energy physics.  One is to go to 
higher energy to search for new particles and new interactions 
and the other is to construct facilities with intense beams and 
search for rare processes or processes which are forbidden within 
the SM.  Kaon and muon rare processes have been contributing
in the latter way.  This will continue to be true because there 
are many future plans including AGS 2000 at BNL, experiments at 
Fermilab Main Injector and Japan Hadron Facility (JHF).  In this talk 
I would like to cover three topics on kaon and muon decays which
are considered to be important in future experiments. Namely,  
(1)$K_L \rightarrow \pi^0\nu\bar{\nu}$
and $K^+ \rightarrow \pi^+\nu\bar{\nu}$,
(2) T violation in $K^+ \rightarrow \pi^0\mu^+ \nu$,
(3) lepton flavor violation(LFV) in muon decays.  
I would like to clarify how these processes are important to explore
physics beyond the SM.
        
\section{$K \rightarrow \pi\nu \bar{\nu}$ and CKM Physics}

In the SM various flavor changing neutral current (FCNC) and 
CP violation processes should be consistently explained by the quark flavor 
mixing matrix called the Cabibbo-Kobayashi-Maskawa (CKM) matrix.
The CKM matrix is parametrized by four independent parameters.
In the Wolfenstein parametrization this matrix is given by
\begin{equation}
\left( \begin{array}{ccc}
        1-\frac{1}{2}\lambda^2 & \lambda& A\lambda^3 (\rho-i\eta)\\
        -\lambda &  1-\frac{1}{2}\lambda^2 & A\lambda^2\\
        A\lambda^3 (1-\rho-i\eta)& -A\lambda^2 & 1
                \end{array} \right).
\end{equation}
The two of the four parameters, $\lambda$ and A, are already known 
well. $\lambda$ corresponds to the Cabibbo mixing and is given by
$\lambda = 0.221\pm0.002$ and A is determined from the inclusive
and exclusive decay of B meson and given by $A = 0.82\pm0.06$.\cite{PDG}
The remaining two parameters, $\rho$ and $\eta$, have not been well
constrained.  The main purpose of flavor physics in B and K
meson decays is to measure various quantities which depend on the
$\rho$  and $\eta$ and put as many constraints as possible on the
unitarity triangle defined in the figure 1.  This can be a new window
to physics beyond the SM.
\begin{figure}
\begin{center}
\mbox{\psfig{figure=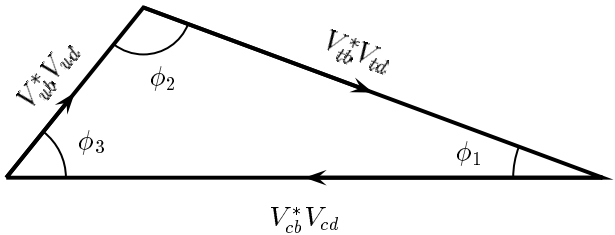,width=4in,angle=0}}
\end{center}
\caption{The unitarity triangle.
\label{fig:fig1}}
\end{figure}

Present constraints on the $\rho$  and $\eta$ parameters are given by three
independent measurements, namely charmless b decay for the ratio of 
the CKM element $|V_{ub}/V_{cb}|$, CP violating mixing parameter
in the $K^0 - \bar{K}^0$ system, $\epsilon_K$, and $B^0_d - \bar{B^0}_d$
mixing.
Although each of three measurements still contains considerable 
theoretical ambiguities from hadron physics, it is remarkable that
there is an overlapping region of the parameter space as shown in figure 2.
\begin{figure}
\begin{center}
\mbox{\psfig{figure=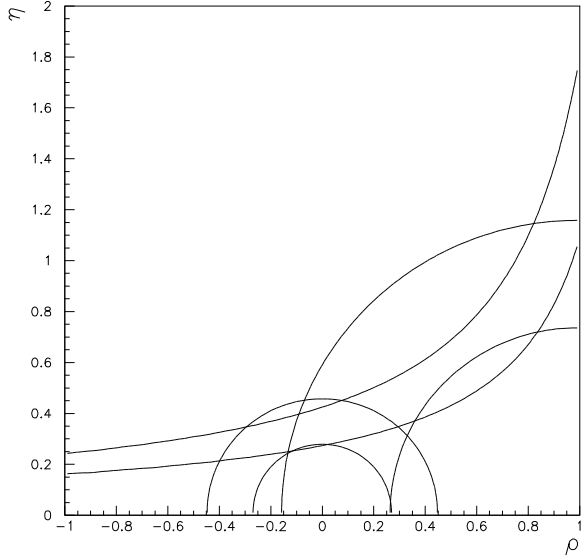,width=3in,angle=0}}
\end{center}
\caption{The constraint on $\rho$  and $\eta$ parameters from
$|V_{ub}/V_{cb}|$, $\epsilon_K$ and $B^0_d - \bar{B^0}_d$.
We take $|V_{ub}/V_{cb}|= 0.08\pm 0.02$, $B_K=0.75\pm 0.15$
and $f_B \surd B_B =200 \pm 40$ MeV.
\label{fig:fig2}}
\end{figure}

In near future, experiments at KEK and SLAC asymmetric B factories 
as well as HERA and TEVATRON will provide us one angle of the unitarity
triangle $\sin2\phi_1$, through the gold-plated mode
$B \rightarrow J/\psi K_S$.
The experimental uncertainty in each of these experiments are expected
to be about 0.1 or smaller for the determination of $\sin2\phi_1$, and
eventually with the LHC-B experiment the precision for the $\sin2\phi_1$
will be a few $\%$ level.  There are many proposals to 
measure other angles of the unitarity triangle.\cite{angle}
The angle $\phi_2$ is determined by time dependent asymmetry of the
$B \rightarrow \pi\pi, \pi\rho$ mode.  The $\phi_3$ measurement can
be done through $B \rightarrow D K$ mode.  Because of small branching
ratios, the determination of these angles requires more luminosity in
the B factory experiments. Another promising way to determine 
the $\rho, \eta$ parameter is measurement of the $B_s - \bar{B}_s$ mixing. 
If we take the ratio of $B_s - \bar{B}_s$ and $B_d - \bar{B}_d$ mixing 
we get
\begin{equation}
\frac{\Delta M_{B_s}}{\Delta M_{B_d}}
=\frac{M_{B_s} B_{B_s} f_{B_s}^2}
{M_{B_d} B_{B_d} f_{B_d}^2}
\left| \frac{V_{ts}}{V_{td}}\right| ^2
\end{equation}
and the ratio of the CKM matrix element depends on the parameter 
$(1-\rho)^2 + \eta^2$.  The bag parameter $B_B$ and the decay constant
$f_B$ should be determined from lattice gauge theory.  
The error of the lattice determination for $B_B f_B^2$ is supposed to be 
smaller if we take the ratio between these quantities for $B_s^0$ and 
$B_d^0$ instead of considering $B f_B^2$ itself.  In the SM the present 
allowed region on ($\rho, \eta$)space predicts the $B_s$ mixing parameter
$x_s\equiv \frac{\Delta M_{B_s}}{\Gamma _{B_s}}$
in the range of $15\lsim x_s \lsim 40$. The HERA-B experiment may be able to
measure the $B_s$ mixing in future and eventually the experiment at LHC will 
be able to cover whole parameter region of $x_s$.\cite{LHC}

Branching ratios of $K^+ \rightarrow \pi^+\nu\bar{\nu}$ and 
$K_L \rightarrow \pi^0\nu\bar{\nu}$ are also important to determine
the $\rho$ and $\eta$ parameters.
Advantages of these processes are theoretical cleanness:
The form factors are determined from the $K^+ \rightarrow \pi^0e^+\nu$
decay.  Also the perturbative QCD corrections are calculated up to 
the next-to-leading order and remaining theoretical ambiguities are 
estimated to be less than 10$\%$ for $K^+ \rightarrow \pi^+\nu\bar{\nu}$ and 
a few $\%$ for $K_L \rightarrow \pi^0\nu\bar{\nu}$.\cite{BBL}  The long
distance contributions are considered to be small.  In the SM
these two processes are induced by Z-penguin and W-box diagrams.
The branching ratio are given by
$Br(K^+ \rightarrow \pi^+\nu\bar{\nu})\simeq 4.2\times 10^{-11}
( \eta^2+(1.4-\rho)^2),  
Br(K_L \rightarrow \pi^0\nu\bar{\nu})\simeq 
1.8\times 10^{-10}\eta^2$.
The experiment at BNL reported one candidate event for 
$K^+ \rightarrow \pi^+\nu\bar{\nu}$ and they obtain
$Br(K^+ \rightarrow \pi^+\nu\bar{\nu})=4.2^{+9.7}_{-3.5} \times
10^{-10}$.\cite{E787} Although the central value is a few times higher 
this results are consistent with the SM prediction.
The present upper bound for $Br(K_L \rightarrow \pi^0\nu\bar{\nu})$
is $1.8 \times 10^{-6}$ from the KTeV experiment\cite{KTeVkpnn} 
which is still 5 orders of magnitudes above the expected region.
Future plans for dedicated experiments are considered for this 
process.\cite{kpnnprop}

Let us discuss how new physics effects can be explored by above
observables in B and K physics.  In this respect it is important
to improve various measurements in both B and K physics.
One way to look for new physics effects is to try to determine
the $\rho$ and $\eta$ parameters by each observable assuming the 
SM.  New physics effects may appear as inconsistency of the 
parameter determinations. We may first observe the time-dependent 
asymmetry of the $B \rightarrow J/\Psi K_S$ mode and use it as 
an input parameter. Then we can determine  $\rho$ and $\eta$
with one more input parameter which can be provided by  
CP asymmetry of B decay from other modes ($B \rightarrow \pi\pi, \pi\rho$,
$B \rightarrow D K$), $\Delta M_{B_s}$ / $\Delta M_{B_d}$
and the branching ratios for $K \rightarrow \pi\nu\bar{\nu}$.
If the $\rho$ and $\eta$ values determined by these observables
do not coincide we can get an important clue to new physics
from a pattern of the deviation from the SM among 
these observables.

One such example is given by the minimal supersymmetric (SUSY)
standard model based on supergravity theory.  In SUSY model
flavor physics is affected by loop diagrams of SUSY particles as
well as the charged Higgs boson and FCNC
processes such as $\epsilon_K$, $\Delta M_B$, 
$Br(K \rightarrow\pi\nu\bar{\nu})$ 
receive new contributions.  Also the CP asymmetry in
B decay may be affected by a new phase of the $B_d^0 - \bar{B}_d^0$
mixing amplitude. In the context of the minimal supergravity model, 
however, we can show that loop contributions to various FCNC amplitudes
associated with SUSY particles and the charged Higgs boson depend on 
essentially the same CKM parameters as the SM contribution.
Thus SUSY contributions can change magnitudes of FCNC amplitudes,
but do not introduce new phases in the $B_d^0 - \bar{B}_d^0$
mixing amplitude. We have calculated in details the $\epsilon_K$ 
and $\Delta M_B$ in the this model.\cite{GNO}  We can also deduce 
$K^+ \rightarrow \pi^+\nu\bar{\nu}$ or
$K_L \rightarrow \pi^0\nu\bar{\nu}$ branching ratio 
for the SUSY model from the calculation of
$Br(b \rightarrow s\nu\bar{\nu})$ \cite{GOST} because these quantities 
are practically the same in the minimal supergravity model 
if normalized by the SM prediction.  
We see that the $\epsilon_K$ and  $\Delta M_B$ are enhanced up 
to 20$\%$ in the minimal supergravity model compared to the SM values 
whereas the branching ratio for $K^+ \rightarrow \pi^+\nu\bar{\nu}$ 
are suppressed up to 5$\%$.  One the other hand the $\rho$ and $\eta$
values determined from the CP asymmetry in B decay and 
$\Delta M_{B_S}$/$\Delta M_{B_d}$
are not affected by the SUSY contributions.  
If we relax the strict universality condition for the SUSY breaking 
terms for Higgs fields and squark fields the deviations from the SM 
can be twice as large as above.\cite{GOS}  In this model, therefore,
we may see difference of ($\rho, \eta$)values determined from 
(1) $\epsilon_K$, $\Delta M_B$,
(2) $Br(K^+ \rightarrow \pi^+\nu\bar{\nu})$,
    $Br(K_L \rightarrow \pi^+\nu\bar{\nu})$ and 
(3) CP asymmetries in $B\rightarrow\pi\pi$ and  
$B\rightarrow DK$ models and 
$\Delta M_{B_S}$/$\Delta M_{B_d}$.
This example suggests importance of measuring the
$Br(K^+ \rightarrow \pi^+\nu\bar{\nu})$ or
$Br(K_L \rightarrow \pi^0\nu\bar{\nu})$ 
at the level of 10$\%$ in order to be competitive 
to other measurements 

\section{T Violation in $K^+ \rightarrow \pi^0\mu^+\nu$ decay}

In the three body decay with initial or final polarization we can 
define a triple vector correlation which is a T odd quantity. 
Assuming CPT invariance this is another way to look for
CP violation. 
In the $K^+ \rightarrow \pi^0\mu^+\nu$ decay the transverse muon 
polarization is a signal of T violation.
The previous limit on this quantities  is 
$P_{\perp}(K^+ \rightarrow \pi^0\mu^+\nu) 
=(-3.1\pm5.3)\times10^{-3}$.\cite{Blatt}
There is an on-going experiment at KEK aiming to improve the 
limit by one order of magnitude. In the SM the 
contribution induced from Kobayashi-Maskawa phase is negligible
and the T odd asymmetry induced by the final state interaction
which mimics the T violation is estimated to be $0(10^{-6})$.\cite{FSI}
Thus this process provides an interesting window to look
for new sources of CP violation in such models as multi-Higgs
doublet model(MHDM)\cite{Kpimunu} and SUSY models.\cite{WN}

In the MHDM without tree-level FCNC 
we can introduce new CP violating phases in the charged Higgs mass
matrix if the number of Higgs doublets is more than three.  The coupling
with charged Higgs boson and quarks is written as 

\begin{equation}
{\cal L }= (2\sqrt{2}G_F)^{\frac{1}{2}}\sum_{i=1}^{2} \{
\alpha_i \bar{u}_L V M_D d_R H_i^+ + \beta_i \bar{u_R} M_U V
d_L H_i^+ + \gamma_i \bar{\nu}_L M_E e_R H_i^+ \} + h.c.,
\end{equation} 
where $M_D, M_U, M_E$ are diagonal mass matrixes and 
$V$ is the CKM matrix.  $\alpha_i, \beta_i, \gamma_i$ are new
complex coupling constants associated to the charged Higgs interactions.
There are several relations among these coupling constants from the 
unitarity of the charged Higgs boson mixing matrix.  Using these
relations we can show that the number of the physical phases is 
$\frac{(n-2)(n-1)}{2}$ for n-Higgs doublet model.

The transverse muon polarization in the
$K^+ \rightarrow \pi^0\mu^+\nu$ decay is induced from the
interference between the W-boson and the charged Higgs boson
exchange diagrams. These diagrams generate operators
of $\bar{s}_L\gamma^{\mu}u_{L}\bar{\nu}_L \gamma_{\mu}\mu_{L}$ and 
$\gamma\alpha^*\bar{s}_{R}u_{L}\bar{\nu}_{L}\mu_{R}$, respectively.
The polarization is therefore proportional to $Im(\gamma\alpha^*)$.
If we assume that only the lightest charged Higgs boson's contribution
is dominant, we get
\begin{equation}
P_{\perp}(K^+ \rightarrow \pi^0\mu^+\nu) 
\simeq -0.2 \frac{m_K^2}{m_H^2}Im(\gamma_1\alpha_1^*),
\end{equation}
where $\alpha_1, \gamma_1$ stand for the coupling constant for the 
lightest charged Higgs boson.  The most severe constraint on this
combination of coupling constants come from the tauonic B decay
which is given by $|Im(\gamma\alpha^*)|\lsim 0.2 
(\frac{m_H}{GeV})^2$.\cite{newbound} 
This leads to the bound on the transverse muon 
polarization as $|P_{\perp}(K^+ \rightarrow \pi^0\mu^+\nu)|
\lsim 1 \times 10^{-2}$ ,which is about the same as the present 
experimental bound.  This means that non-zero value of the transverse 
polarization may be observable at the on-going experiment and 
if not the improved bound puts the strongest constraint on this 
combination of the coupling constants.

The transverse muon polarization in the 
$K^+ \rightarrow \mu^+\nu\gamma$ process is also a signal of
T violation. Although the branching ratio is smaller, this process
can be measured in the same on-going KEK experiment.
The effect of final state interaction is estimated to be $0(10^{-3})$
\cite{Geng} so that this effect has to be subtracted properly 
in future experiments. It is interesting to see correlation between 
two transverse polarizations. In the MHDM, we obtain
\begin{equation}
P_{\perp}(K^+ \rightarrow \mu^+\nu\gamma) 
\simeq -0.1 \frac{m_K^2}{m_H^2}Im(\gamma_1\alpha_1^*),
\end{equation}
if we use various phenomenological constraints on $\alpha, \beta, \gamma$
coupling constants.\cite{KLO}  Therefore if these polarizations are observed 
in near future, two polarizations should be in the same sign.

There are many models which induce transverse muon polarization at 
the observable level in future experiments. One interesting example
is the SUSY model discussed by G.-H. Wu and J.N. Ng.\cite{WN}  
In this case a complex coupling constant between the charged Higgs 
boson and strange and up quarks
can be induced through scalar quark and gluino loops.  
Although this is a loop effect it is possible to find a 
parameter region where the induced transverse polarization for 
$K^+ \rightarrow \pi^0\mu^+\nu$ is
$0(10^{-3})$ if we allow large flavor mixing coupling in 
squawk-quark-gluino vertices and take a large value for the ratio
of two vacuum expectation values($\tan\beta$).  Also the 
correlation between 
$P_{\perp}(K^+ \rightarrow \pi^0\mu^+\nu)$ and 
$P_{\perp}(K^+ \rightarrow \gamma\mu^+\nu)$ 
is opposite in sign if the transverse 
polarization is induced by this loop effect.\cite{WNgamma} 
It is therefore useful to search for transverse muon polarization
in both processes in order to investigate nature of new CP violation
interactions.

\section{Lepton flavor violation in muon decay}
In the minimal SM, electron, muon and tau lepton numbers are separately
conserved and there is no lepton flavor violation (LFV).
This property is associated with the fact that we  cannot write 
gauge-invariant and renormalizable interactions with LFV within the
SM. If we introduce extra fields or interactions we can easily break 
conservation of separate lepton numbers. Although there are many 
experimental searches for LFV in $\tau$,  $Z^0$ and K decays, muon rare
decays put particularly strong constraints. Experimental upper bounds
quoted in PDG 96 are $4.9 \times 10^{-11}$ for the $\mu^+ \rightarrow
e^+ \gamma$branching ratio, $1.0 \times 10^{-12}$ for the $\mu^+ \rightarrow
e^+ e^+ e^-$branching ratio and $4.3 \times 10^{-12}$ for the  
$\mu^-  - e^-$ conversion rate in T$_i$ atoms normalized to the muon capture
rate.\cite{PDG} 
For the $\mu^+ \rightarrow e^+ \gamma$ process the MEGA experiment
at Los Alamos National Laboratory is analysing data and aiming to improve 
the branching ratio by one order of magnitude. The $\mu^- - e^-$ conversion 
experiment is also continued at Paul Scherrer Institute to search for
this process at the level of $O(10^{-14})$.

Among various models which predict sizable LFV, SUSY models attract much
attention recently. In particular it has been pointed out that SUSY
GUT models can induce LFV at the level close to the present experimental
bounds. In some cases a part of the SUSY parameter space is already 
excluded by the LFV processes, especially in SO(10) SUSY GUT model.
\cite{BH,BHS}

Unlike the minimal SM, the SUSY SM does not necessarily conserve the lepton 
number separately for each generation. This is because that the mass 
matrices for scalar partner of leptons, {\it i.e.} sleptons, can be a new
source of flavor mixing in addition to the Yukawa coupling constants. 
Although these mass terms conserve total lepton number, they do not have 
to conserve lepton number for each generation separately. Since the scalar
mass terms are determined by SUSY breaking terms the LFV depends on
how SUSY is broken spontaneously at the energy scale higher than the 
electroweak scale. In fact if we allow arbitrary flavor mixing
in the slepton sector, too large LFV is often induced. A similar problem
occurs in the squark sector where the $K^0 -\bar{K} ^0$ mixing becomes
too large unless some suppression mechanism is implemented. In the minimal
supergravity model these flavor problems can be avoided because SUSY 
breaking masses for all scalar fields are assumed to be universal at the 
Planck scale. LFV processes is therefore forbidden in this model 
if there is no LFV interaction between the Planck and
electroweak scales. On the other hand LFV processes are induced through 
renormalization effects on slepton mass 
matrices if some LFV interaction is present below the Planck scale.

In the SU(5) SUSY GUT model the LFV mass term for the right-handed slepton
is induced through renormalization effects between the Planck and GUT 
scales. Above the GUT scale the right-handed slepton mass terms receive
a loop correction from the top Yukawa coupling constant because the  
right-handed slepton is included in the {\bf 10} dimensional 
representation of SU(5) and the top Yukawa coupling  is in the form of     
${\bf 10}\cdot {\bf 10}\cdot {\bf H(5)}$  where ${\bf H(5)}$ represents
the {\bf 5} dimensional representation Higgs field. Due to this 
renormalization effect the third generation right-handed slepton,
{\it i.e.} the right-handed stau becomes lighter than other two 
right-handed sleptons and the slepton mass matrix is no longer 
proportional to a unit matrix. Thus this matrix cannot be diagonalized 
simultaneously with the lepton mass matrix. In the paper by Barbieri 
and Hall it is pointed out that the induced $\mu \rightarrow e \gamma$
branching ratio can be close to the present experimental upper bound
mainly due to the effect of large top Yukawa coupling constant\cite{BH}.
Precise value of the branching ratio depends on various SUSY parameters
as well as assumption on Yukawa coupling constants at the GUT scale.
According to the recent detailed calculation in this model, the  
$\mu \rightarrow e \gamma$ branching ratio can be as large as $10^{-13}$
especially for large values of $\tan{\beta}$ if we make a simple 
assumption that Yukawa coupling constant are solely given by    
${\bf 10}\cdot {\bf 10}\cdot {\bf H(5)}$ and ${\bf 10}\cdot {\bf \bar{5}}
\cdot {\bf H(\bar{5})}$ couplings at the GUT scale.\cite{HMTY2}

In the SO(10) model dominant contribution to the LFV amplitude is 
given by diagrams proportional to tau-lepton mass in the slepton 
internal line. Compared to the SU(5) case the branching ratio can 
be enhanced by $(\frac{m_{\tau}}{m_{\mu}})^2$ and therefore a 
large part of SUSY parameter space is already 
excluded by the $\mu \rightarrow e \gamma$ process.\cite{BHS} 
In this model we can derive approximate relations among rates 
of $\mu^+ \rightarrow e^+ \gamma$, $\mu^-  - e^-$ conversion 
and $\mu^+ \rightarrow e^+ e^+ e^-$ processes. Namely, 
\begin {eqnarray} 
\frac{\sigma(\mu T_i \rightarrow e T_i)}
{\sigma(\mu T_i \rightarrow capture)}
&\simeq & \frac{1}{200} Br(\mu^+ \rightarrow e^+ \gamma),\\
Br(\mu^+ \rightarrow e^+ e^+ e^-)
&\simeq & \frac{1}{150} Br(\mu^+ \rightarrow e^+ \gamma)
\end {eqnarray}
hold. Although the $\mu^-  - e^-$ conversion  and $\mu^+ \rightarrow
e^+ e^+ e^-$ processes depend on four-fermion operators
in addition to the photon penguin operator, only the latter one 
receives enhanced contributions from the above mentioned
diagram. 

Large LFV may be induced in the SUSY model with small neutrino mass
induced by see-saw mechanism.\cite{HMTY}
In this case the Yukawa coupling constant 
for right-handed neutrino superfield and Majorana mass terms for
right-handed neutrinos can be new sources of LFV and if the Yukawa 
coupling constant is large enough the left-handed slepton mass terms
receive generation dependent corrections. If we use the see-saw relation 
for the neutrino mass as $m_{\nu}\sim \frac{m_D^2}{M_N}$ where $m_D$ is the 
Dirac mass and $M_N$ is the Majorana mass for right-handed neutrino, the 
Yukawa coupling constant becomes as large as the top Yukawa coupling constant 
for $M_N \sim 10^{13}$ GeV and $m_{\nu}\sim 1$ eV. 
Since the $\mu \rightarrow e \gamma$ 
branching ratio strongly depends on unknown parameters such as
the right-handed Majorana mass scale and mixing matrix elements 
the prediction for the branching ratio is more ambiguous in this model, 
but it is interesting to see that in large fraction of parameter space  
LFV effects are large enough to be observed in near-future experiments. 
This is contrasted to the see-saw neutrino model without SUSY where the 
branching ratio for $\mu \rightarrow e \gamma$ etc. are too small for 
experiments in near future although evidence of LFV may be obtained through 
neutrino oscillation experiments.

Let us finally comment on usefulness of muon polarization
in search for LFV. A highly polarized muon beam is available in 
$\mu^+$ decay experiments. Muons from $\pi^+$ decay stopped near 
the surface of the pion production target is 100\% polarized opposite
to the muon momentum and this muon is called surface muon.

The first obvious merit of polarized muons in 
$\mu^+ \rightarrow e^+ \gamma$ is that we can distinguish   
$\mu^+ \rightarrow e^+_R \gamma$ and 
$\mu^+ \rightarrow e^+_L \gamma$ by the angular distribution
of the decay products with respect to the muon polarization direction.
For example, the positron from the $\mu^+ \rightarrow e^+_R \gamma$
decay follows the $(1-P \cos{\theta})$ distribution where $\theta$ 
is the angle between the polarization direction and the positron
momentum and $P$ is the the muon polarization. In the previous examples
the SU(5) SUSY GUT predicts $\mu^+ \rightarrow e^+_L \gamma$
because LFV is induced only in the right-handed slepton sector.
On the other hand the SO(10) SUSY GUT generates almost equal number
of $\mu^+ \rightarrow e^+_L \gamma$ and $\mu^+ \rightarrow e^+_R \gamma$
so that the positron has a flat angular distribution. If the LFV
is induced by the right-handed neutrino Yukawa coupling constant,
only $\mu^+ \rightarrow e^+_R \gamma$ should be observed.

Polarized muons are also useful to suppress background processes 
for the $\mu^+ \rightarrow e^+ \gamma$ search.\cite{polmu} 
In this experiment the experimental sensitivity is limited by 
appearance of the background processes. There are two major background
processes. The first one is physics background process which is a tail 
of radiative muon decay. If neutrino pair carries out only little energy
in the $\mu^+ \rightarrow e^+ \nu \bar{\nu} \gamma$ process, we cannot
distinguish this from the signal process. The second background process 
is an accidental background process where detections of 52 MeV positron
and 52 MeV photon from different muon decays coincide within time
and angular resolutions for selection of signals. The source of the  
52 MeV positron is the ordinary $\mu^+ \rightarrow e^+ \nu \bar{\nu}$
decay whereas the 52 MeV photon  mainly comes from a tail of
the radiative muon decay. We calculated the angular distribution
of the final positron and photon and showed that polarized muons are 
useful for suppression of both background processes. For the physics 
background it can be shown that the positron follows approximately
$(1+P \cos{\theta})$ distribution if we take into account finite
energy resolution of photon and positron detectors. The physics 
background is therefore suppressed for the
$\mu^+ \rightarrow e^+_R \gamma$ search if the polarized muon is
used. For the accidental background both positrons and photons turn
out to follow $(1+P \cos{\theta})$ distribution. Thus background 
suppression works independently of the signal distribution.
If we use 97\% polarized muons we can expect to reduce the accidental
background by one order of magnitude. This looks promising for search
of $\mu^+ \rightarrow e^+ \gamma$ at the level of $10^{-14}$ branching 
ratio.

The third example of the merit of polarized muon decays is
that we can measure T and CP violation in the   
$\mu^+ \rightarrow e^+ e^+ e^-$ decay.\cite{m3eCP} Since we can take a
triple vector correlation for three body decays of polarized 
particles, T odd asymmetry can be defined in 
$\mu^+ \rightarrow e^+ e^+ e^-$ decay. We calculated that 
T odd asymmetry in SU(5) SUSY GUT and showed that the asymmetry can
be as large as 20\% if we include CP violating phases in SUSY
soft breaking terms.\cite{OOS} If LFV is discovered in the 
$\mu^+ \rightarrow e^+ \gamma$ or $\mu ^+ \rightarrow e^+ e^+ e^-$
processes measurement of T odd asymmetry will become an important
next target which could provide us information on CP nature
of LFV interactions.
  
\section{Conclusions}
I have reviewed three topics on kaon and muon decays: (1) $K_L \rightarrow 
\pi^0\nu\bar{\nu}$ and $K^+ \rightarrow \pi^+\nu\bar{\nu}$,
(2) T violation in $K^+ \rightarrow \pi^0\mu^+ \nu$,
(3) LFV in muon decays. 
These precesses can be new windows to physics beyond the SM.
Measurement of the branching ratio of $K_L \rightarrow \pi^0\nu\bar{\nu}$ and/or
$K^+ \rightarrow \pi^+\nu\bar{\nu}$ at the 10\% level will provide us   
an important clue for new physics if we combine with other observables
related to the CKM matrix in B decays. We can obtain information
of new CP sources in MHDM and SUSY models from the measurement of transverse 
muon polarization in $K^+ \rightarrow \pi^0\mu^+ \nu$. The 
correlation with transverse muon polarization in $K^+ \rightarrow 
\mu^+\nu\gamma$ is useful to distinguish various models.
Finally  we showed that LFV processes such as $\mu^+ \rightarrow e^+ \gamma$,
$\mu ^+ \rightarrow e^+ e^+ e^-$ and $\mu^-  - e^-$ conversion 
in atoms can occur at the rate as large as the present experimental 
upper bounds in SUSY GUT and the SUSY model with right-handed neutrino.
If observed, these processes can give us information on LFV interaction 
at very high energy scale.  These experiments therefore may provide 
us the first hint for the physics beyond the SM in facilities with an 
intense proton beam such as JHF.
\newpage


%

\section{References}


\newcommand{\Journal}[4]{{#1} {\bf #2}, {(#3)} {#4}}
\newcommand{\pl}{\sl Phys.~Lett.}
\newcommand{\plb}{\sl Phys.~Lett.~{\bf B}}
\newcommand{\prp}{\sl Phys.~Rep.}
\newcommand{\pr}{\sl Phys.~Rev.}
\newcommand{\prd}{\sl Phys.~Rev.~{\bf D}}
\newcommand{\prl}{\sl Phys.~Rev.~Lett.}
\newcommand{\np}{\sl Nucl.~Phys.}
\newcommand{\npb}{\sl Nucl.~Phys.~{\bf B}}
\newcommand{\ptp}{\sl Prog.~Theor.~Phys.}
\newcommand{\zp}{\sl Z.~Phys.}
\newcommand{\zpc}{\sl Z.~Phys.~{\bf C}}
\newcommand{\mpl}{\sl Mod.~Phys.~Lett.}
\newcommand{\rmp}{\sl Rev.~Mod.~Phys.}
\newcommand{\mpla}{\sl Mod.~Phys.~Lett.~{\bf A}}
\newcommand{\sjnp}{\sl Sov.~J.~Nucl.~Phys.}
\newcommand{\ibid}{\it ibid.}


\end{document}